# A MODEL OF HOMOGENEOUS SEMICOHERENT INTERPHASE BOUNDARY FOR HETEROPHASE PRECIPITATES IN SUBSTITUTION ALLOYS UNDER IRRADIATION

A. Borisenko

*National Science Centre «Kharkiv Institute of Physics and Technology», Akademichna St. 1, 61108 Kharkiv, Ukraine;*

*e-mail: borisenko@kipt.kharkov.ua*

*tel.: +380 57 3356203*

*fax: +380 57 3352683*

**Abstract**

The model of homogeneous semicoherent interphase boundary describes the processes of absorption and thermoactivated migration of irradiation-produced inequilibrium point defects at a semicoherent boundary between a heterophase precipitate and a substitution solid solution. Within this model the kinetics of evolution of the sizes of precipitates of constant chemical composition under irradiation is investigated. The results obtained are compared to the experimental data [I. Monnet et al., J. Nucl. Mater. 335 (2004) 311] for the ferritic ODS steel EM10+MgO under electron irradiation.

## 1. Introduction

There are several mechanisms of irradiation influence on stability of heterophase precipitates[1] in alloys (see e.g. [1]). Radiation disordering violates the phase stability of stochiometric precipitates. Ballistic effects lead to the forced transport of substance between precipitate and matrix. If precipitates contain sinks of point defects, then the radiation-induced segregation of alloy's components in their vicinities takes place (the inverse Kirkendall effect). Irradiation can cause polymorphous transformations (martensite transitions, amorphization, etc.). Migration of inequilibrium point defects between a precipitate and a matrix can lead to change of a chemical composition of the precipitate. Irradiation-produced continuous defects of crystalline structure (dislocation loops, pores, etc.) can be nucleation centres for secondary precipitates.

However, some recent experimental data have no explanation in the framework of existing theoretical models. In the paper [2] the complex analysis of the effect of different types of irradiation on stability of precipitates in several types of ODS ferritic martensitic steels is carried out. It has been shown, that neutron irradiation in Phénix reactor leads to dissolution of $Y_2O_3$–based precipitates at high temperatures and irradiation doses. To reveal the mechanisms of precipitate dissolution, different irradiations by several types of charged particles were performed. It turned out that irradiation by helium ions with energy 1 MeV (the region of domination of electron losses, where the point defects are not produced) has no effect on the precipitates in the steels studied. Irradiation by argon ions with energy 300 keV, which produce cascades of point defects (PD), leads to dissolution of the precipitates. Irradiation by 1 MeV and 1.2 MeV electrons, which produce isolated PD, leads to considerable decrease of the precipitates based on MgO and $Y_2O_3$ respectively. Thus it was demonstrated, that these are isolated PD which are basically responsible for precipitate dissolution. Electron irradiation with energy 1 MeV does not affect $Y_2O_3$-based precipitates, and 1.2 MeV electron irradiation leads to their dissolution. The displacement threshold energy for yttrium ion in the $Y_2O_3$ lattice is $E_d = 57$ eV [2]. During electron-yttrium collision, the maximum energy transferred is $E_t = 49$ eV by 1 MeV electron and $E_t = 64$ eV by 1.2 MeV electron [3]. Thus, the necessary condition of precipitate dissolution is production of PD inside the precipitate. Therefore, the mechanism of radiation-induced segregation [4] in this case does not play a crucial role.

---

[1] Hereinafter we consider the heterophase precipitates which are different from the matrix alloy in their chemical composition.

The analysis [2] of experimental data on dissolution of MgO-based precipitates under 1 MeV electron irradiation shows, that the ballistic mechanism of dissolution [5, 6] yields about 10 % of the observable velocity of precipitate dissolution. Besides, the observable velocity of precipitate dissolution increases with temperature increase, whereas in both cases of radiation disordering relaxation and back-diffusion of recoil atoms one expects for an opposite temperature effect.

This paper presents a microscopic model of substance transport between a heterophase precipitate (below – a precipitate) and a substitution solid solution (matrix) due to thermally activated migration across the interphase boundary of inequilibrium PD produced by irradiation. Consider a precipitate with concentration of one of its components, conventionally designated A (below – an impurity) $c_A^p = const$, located in a matrix with impurity concentration $c_A^m$, such that $c_A^p > c_A^m$. Irradiation produces inequilibrium PD in the precipitate and in the matrix. Since concentration of the impurity interstitials in the precipitate is greater than in the matrix, their diffusion flux through the interphase boundary is directed from the precipitate to the matrix. If concentration of the impurity in lattice sites of the matrix near the boundary exceeds its solubility limit in the solid solution, then there is an opposite flux of the impurity from the matrix to the precipitate by a vacancy mechanism. The competition of these two fluxes determines the kinetics of evolution of the precipitate size.

## 2. A model of homogeneous semicoherent interphase boundary

Owing to the misfit of crystalline lattices of the adjacent phases the shear stresses occur at the interphase boundary. These stresses increase with increasing of the precipitate size and at reaching of some threshold value they are compensated by formation of a network of misfit dislocations at the boundary surface. Such a boundary is conventionally called semicoherent. The cores of misfit dislocations are sinks for inequilibrium PD. At the same time, the coherent regions of the boundary are transparent for PD. Therefore, the flux of PD of the type $n$ ($n = i$ – interstitial, $n = v$ – vacancy), averaged over the boundary surface, in a phase $\varphi$ ($\varphi = p$ – precipitate, $\varphi = m$ – matrix) near the boundary can be represented as a sum of the absorbed $j_{n\,a}^{\varphi}(r_p)$ and transited $j_{n t}(r_p)$ components:

$$j_n^{\varphi}(r_p) = j_{n\,a}^{\varphi}(r_p) + j_{n t}(r_p). \tag{1}$$

Below in the sections 2.1 and 2.2 within the framework of the model of homogeneous semicoherent interphase boundary we find an explicit form for $j_{n\,a}^{\varphi}(r_p)$ and $j_{n t}(r_p)$, and in the section 2.3 we derive a kinetic equilibrium impurity concentration under irradiation conditions.

The stationary concentration profiles of PD with diffusion constants $D_n^{\varphi}$ in an effective homogeneous absorbing medium with a volume concentration of sinks $(\kappa^{\varphi})^2$, which is supposed to be identical for vacancies and interstitials, are set up during the characteristic times $\tau_{\kappa} \sim \left[D_n^{\varphi}(\kappa^{\varphi})^2\right]^{-1}$, which are much less than one second already at a room temperature. Therefore, one can consider the PD diffusion to be stationary. In further description the mutual recombination of interstitials and vacancies is neglected. This assumption is valid for not very high PD concentrations, when the characteristic recombination length is greater than the mean distance between PD sinks [1].

### 2.1 Absorption of PD at the interphase boundary

Consider a semicoherent interphase boundary (below – a boundary) containing a network of misfit dislocations with a mean distance $L$ between the dislocation cores with radius $r_d$. For a square network of misfit dislocations their surface density (the length of the dislocation line per surface unity) is given by the expression:

$$\sigma_d = 2/L. \tag{2}$$

To estimate the value of $j_{n\,a}^{\varphi}(r_p)$ we employ the following model. Each misfit dislocation is assigned its region of influence in the form of a half of cylinder with radius $L/2$ in both phases (see Fig. 1).

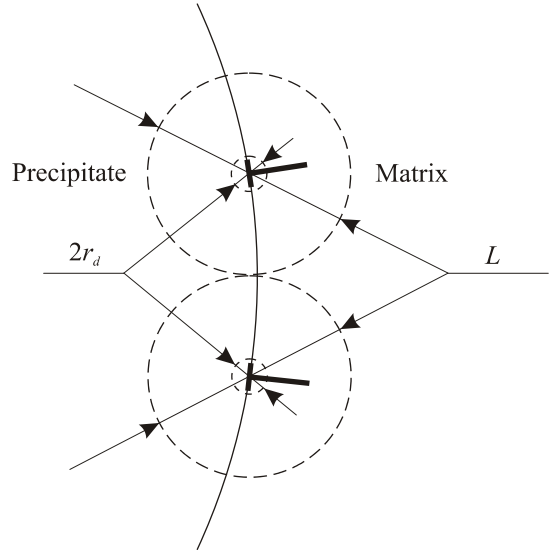

Fig. 1. Regions of influence and cores of misfit dislocations in the vicinity of an interphase boundary.

In this way, the whole surface of the interphase boundary appears to be divided into the regions of influence. Suppose that a region of influence is free from other sinks and sources of PD, and the space out of a region of influence is a homogeneous effective absorbing medium with PD concentration $c_{n\,vol}^{\varphi}$. This assumption is valid for small values of $L$:

$$L << \min\{r_p, l_{rec}, l_{abs}\}, \qquad (3)$$

where $r_p$ is a precipitate radius, and $l_{rec}$ and $l_{abs}$ are respectively recombination and absorption lengths for PD.

Then the diffusion of PD in the region of influence is governed by the following diffusion problem:

$$\mathrm{div}\, j_n^{\varphi} = 0, \quad j_n^{\varphi} = -D_n^{\varphi} \nabla c_n^{\varphi}; \qquad (4)$$

$$c_n^{\varphi}(r_d) = c_n^{\varphi 0}, \qquad (5)$$

$$c_n^{\varphi}(L/2) = c_{n\,vol}^{\varphi}, \qquad (6)$$

where $D_n^{\varphi}$ is a PD diffusion constant, $c_n^{\varphi 0}$ is a thermodynamic equilibrium PD concentration.

The equation (4) with the boundary conditions (5), (6) has the following solution:

$$c_n^{\varphi}(r) = \left(c_{n\,vol}^{\varphi} - c_n^{\varphi 0}\right) \frac{\ln(r/r_d)}{\ln(L/2r_d)} + c_n^{\varphi 0}. \qquad (7)$$

The total PD flux per unity of length of the dislocation line is given by the expression:

$$J_n^{\varphi} = \oint_{S \in \varphi} j_n^{\varphi} ds = \frac{\pi D_n^{\varphi} \left(c_{n\,vol}^{\varphi} - c_n^{\varphi 0}\right)}{\ln(L/2r_d)}. \qquad (8)$$

Then the average PD flux absorbed at the interphase boundary (a unit normal vector is directed to the boundary) is given by the expression:

$$j_{n\,a}^{\varphi}(r_p) = J_n^{\varphi}\sigma_d = \frac{2\pi D_n^{\varphi}\left(c_{n\,vol}^{\varphi} - c_n^{\varphi 0}\right)}{L\ln(L/2r_d)}. \qquad (9)$$

Now we employ the approach of an effective homogeneous absorbing boundary. For this purpose one needs to express the flux of absorbed PD (9) in terms of their mean concentration at the boundary. From (7) one finds the mean concentration of PD inside the region of influence (and, therefore, at the boundary):

$$c_n^{\varphi}(r_p) = \frac{\int_{r_d}^{L/2} c_n^{\varphi}(r)r\,dr}{\frac{1}{2}(L/2)^2} = \left(c_{n\,vol}^{\varphi} - c_n^{\varphi 0}\right)\left\{1 - \frac{1}{2}\left[1 - \left(\frac{2r_d}{L}\right)^2\right]\left[\ln\left(\frac{L}{2r_d}\right)\right]^{-1}\right\} + c_n^{\varphi 0}. \qquad (10)$$

Then from (9) and (10) it is straightforward to find a relation between the flux of PD, absorbed at the boundary, and their mean concentration at the boundary:

$$j_{n\,a}^{\varphi}(r_p) = \alpha_n^{\varphi}\left[c_n^{\varphi}(r_p) - c_n^{\varphi 0}\right], \qquad (11)$$

where

$$\alpha_n^{\varphi} = \frac{2\pi D_n^{\varphi}}{L\left\{\ln(L/2r_d) - \frac{1}{2}\left[1 - (2r_d/L)^2\right]\right\}}. \qquad (12)$$

From (11) and (12) it follows that there is a critical value of the distance between misfit dislocations $L_c = 2r_d$, at which the quantity $\left[c_n^{\varphi}(r_p) - c_n^{\varphi 0}\right]$ turns to zero, i.e. the boundary becomes completely incoherent. The case of $\alpha_n^{\varphi} \to 0$ is realized at $L \to \infty$ and corresponds to a coherent boundary.

**2.2 Transition of PD across the interphase boundary**

Within this model the PD concentration profiles are supposed to be discontinuous at the interphase boundary. Therefore, PD transition across the boundary is considered as a reversible surface chemical reaction.

Thus, a frequency of transitions of interstitials of each type across the boundary is proportional to their partial concentration in the corresponding phase. Therefore, the flux of the interstitials, transited across the boundary, is given by the following expression (hereinafter the normal unit vector is considered to be guided from the precipitate into the matrix):

$$j_{it}(r_p) = \sum_{A \in p}\left[\tilde{\beta}_{iA}^{p} c_{iA}^{p}(r_p) - \tilde{\beta}_{iA}^{m} c_{iA}^{m}(r_p)\right], \qquad (13)$$

where the summation is taken over the precipitate's components.

The kinetic coefficients in (13) depend on temperature according to the Arrhenius's law:

$$\widetilde{\beta}_{iA}^{\varphi} \propto \exp\left(-\frac{G_{\beta_{iA}^{\varphi}}(r_p)}{k_B T}\right), \qquad (14)$$

where $G_{\beta_{iA}^{\varphi}}(r_p)$ is an activation energy of transition of an interstitial of type A in the corresponding phase, $k_B$ is the Boltzmann's constant, $T$ is a temperature.

Within this model the exchange of places between a vacancy and an atom A, being on the different sides of the boundary, is represented in the form of a reversible chemical reaction:

$$l_A^m + v_A^p \leftrightarrow l_A^p + v^m, \qquad (15)$$

where $l_A^m$ is an atom A in the regular site of the matrix lattice; $v_A^p$ is a vacancy in the A sublattice of the precipitate; $l_A^p$ is an atom A in the regular site of the precipitate lattice; $v^m$ is a vacancy in the matrix. Hereinafter it is considered, that atoms in the precipitate can occupy sites only in their native sublattices. In the spirit of the theory of velocities of chemical reactions, the flux of the vacancies, transited across the boundary, is given by the following expression:

$$j_{vt}(r_p) = \sum_{A \in p}\left[\widetilde{\beta}_{vA}^{p} c_{vA}^{p}(r_p) c_A^m(r_p) - \widetilde{\beta}_{vA}^{m} c_v^m(r_p) c_A^p\right], \qquad (16)$$

where the summation is taken over the precipitate's components.

The kinetic coefficients in (16) depend on temperature according to the Arrhenius's law:

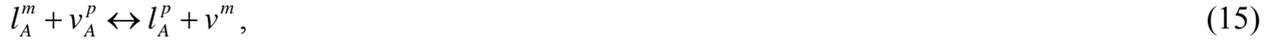

$$\widetilde{\beta}_{vA}^{\varphi} \propto \exp\left(-\frac{G_{\beta_{vA}^{\varphi}}(r_p)}{k_B T}\right), \qquad (17)$$

where $G_{\beta_{vA}^{\varphi}}(r_p)$ is an activation energy of transition of a vacancy to the place of an atom A in the corresponding phase.

The total atomic flux, transited across the boundary, is given by the following expression:

$$j_t(r_p) = j_{it}(r_p) - j_{vt}(r_p). \qquad (18)$$

From the requirement of conservation of a chemical composition of the precipitate, the partial flux of atoms A, transited across the boundary, is related to the total atomic flux (18) by the expression:

$$j_{At}(r_p) = j_t(r_p) c_A^p \omega_0, \qquad (19)$$

where $\omega_0$ is a mean atomic volume.

Consider, that partial concentrations of interstitials and vacancies in the precipitate, and also interstitials in the matrix are related to their total concentrations as follows:

$$c_{nA}^p = x_A^p \omega_0 c_A^p c_n^p, \quad n = i, v; \quad c_{iA}^m = x_A^m \omega_0 c_A^m c_i^m, \qquad (20)$$

where $x_A^\varphi$ is a dimensionless constant, accounting for a possible deviation of PD partial concentrations from the relation $c_{nA}^\varphi / c_n^\varphi = c_A^\varphi \omega_0$.

Then from (13), (16), (18), taking into account (20), one obtains:

$$j_{At}(r_p) = \omega_0 c_A^p [\beta_{iA}^p c_i^p(r_p) + \beta_{vA}^m c_v^m(r_p)] - \omega_0 c_A^m(r_p)[\beta_{iA}^m c_i^m(r_p) + \beta_{vA}^p c_v^p(r_p)], \quad (21)$$

where for brevity of records the kinetic coefficients are renormalized as follows:

$$\beta_{iA}^p = \tilde{\beta}_{iA}^p x_A^p, \quad \beta_{vA}^p = \tilde{\beta}_{vA}^p x_A^p c_A^p, \quad \beta_{iA}^m = \tilde{\beta}_{iA}^m x_A^m, \quad \beta_{vA}^m = \tilde{\beta}_{vA}^m / \omega_0.$$

From comparison of (19) and (21) one obtains:

$$j_t(r_p) = \beta_{iA}^p c_i^p(r_p) + \beta_{vA}^m c_v^m(r_p) - [\beta_{iA}^m c_i^m(r_p) + \beta_{vA}^p c_v^p(r_p)] c_A^m(r_p)/c_A^p, \quad (22)$$

where A is some[2] component of the precipitate.

From comparison of (13), (16), (18), (22) it follows that

$$j_{it}(r_p) = \beta_{iA}^p c_i^p(r_p) - \beta_{iA}^m c_i^m(r_p) c_A^m(r_p)/c_A^p, \quad (23)$$

$$j_{vt}(r_p) = \beta_{vA}^p c_v^p(r_p) c_A^m(r_p)/c_A^p - \beta_{vA}^m c_v^m(r_p). \quad (24)$$

### 2.3. A local kinetic equilibrium at the interphase boundary

The state of the local kinetic equilibrium at the interphase boundary is determined by the requirement that the atomic flux transited across the interphase boundary turns to zero:

$$j_t^{eq}(r_p) = 0. \quad (25)$$

Taking into account (25), one can find from (22) the relation between the equilibrium impurity and PD concentrations at the boundary:

$$\frac{c_A^{m\,eq}(r_p)}{c_A^p} = \frac{\beta_{iA}^p c_i^{p\,eq}(r_p) + \beta_{vA}^m c_v^{m\,eq}(r_p)}{\beta_{iA}^m c_i^{m\,eq}(r_p) + \beta_{vA}^p c_v^{p\,eq}(r_p)}. \quad (26)$$

The equilibrium PD concentrations at both sides of the boundary are governed by the corresponding diffusion equations:

$$\text{div } j_i^p = K^p - D_i^p (\kappa^p)^2 c_i^p, \quad j_i^p = -D_i^p \nabla c_i^p, \quad (27)$$

$$\text{div } j_v^p = K^p - D_v^p (\kappa^p)^2 (c_v^p - c_v^{p0}), \quad j_v^p = -D_v^p \nabla c_v^p, \quad (28)$$

$$\text{div } j_i^m = K^m - D_i^m (\kappa^m)^2 c_i^m, \quad j_i^m = -D_i^m \nabla c_i^m, \quad (29)$$

$$\text{div } j_v^m = K^m - D_v^m (\kappa^m)^2 (c_v^m - c_v^{m0}), \quad j_v^m = -D_v^m \nabla c_v^m, \quad (30)$$

where $K^p$ and $K^m$ are the volume PD generation rates in the precipitate and in the matrix respectively. The first set of boundary conditions for the diffusion equations (27) – (30) is given by the expressions (1), (11), (23), (24) in the following explicit form:

---

[2] The kinetics of evolution of the size of a multicomponent precipitate is usually considered to be determined by the velocity of migration of its slowest component. In this way it is possible to make an unambiguous choice of the component A.

$$j_i^p(r_p) = (\alpha_i^p + \beta_{iA}^p)c_i^p(r_p) - \beta_{iA}^m c_i^m(r_p)c_A^m(r_p)/c_A^p, \tag{31}$$

$$j_v^p(r_p) = \alpha_v^p\left[c_v^p(r_p) - c_v^{p0}\right] + \beta_{vA}^p c_v^p(r_p)c_A^m(r_p)/c_A^p - \beta_{vA}^m c_v^m(r_p), \tag{32}$$

$$j_i^m(r_p) = -\left(\alpha_i^m + \beta_{iA}^m c_A^m(r_p)/c_A^p\right)c_i^m(r_p) + \beta_{iA}^p c_i^p(r_p), \tag{33}$$

$$j_v^m(r_p) = -\alpha_v^m\left[c_v^m(r_p) - c_v^{m0}\right] + \beta_{vA}^p c_v^p(r_p)c_A^m(r_p)/c_A^p - \beta_{vA}^m c_v^m(r_p). \tag{34}$$

The second set of boundary conditions represents the requirement of finiteness of the PD concentrations:

$$c_i^p(0) < \infty, \tag{35}$$

$$c_v^p(0) < \infty, \tag{36}$$

$$c_i^m(\infty) < \infty, \tag{37}$$

$$c_v^m(\infty) < \infty. \tag{38}$$

In the state of the local kinetic equilibrium the impurity and PD concentrations at the boundary are subject to the relation (26). Then the equations (27) – (30) with the boundary conditions (31) – (34) and (35) – (38) represent a diffusion problem.

As a result of its solution one can find a value of the equilibrium impurity concentration at the boundary:

$$\frac{c_A^{m\,eq}(r_p)}{c_A^p} = \frac{a_1}{2a_2}\left(\sqrt{1 - \frac{4a_2 a_0}{a_1^2}} - 1\right), \tag{39}$$

where

$$a_2 = \frac{\beta_{iA}^m}{D_i^m}\frac{\beta_{vA}^p}{D_v^p}\left[c_v^{p0} D_v^p\left(r_p^{-1} - \kappa^p \coth(\kappa^p r_p) - \frac{\alpha_i^p}{D_i^p}\right) + \frac{K^p}{(\kappa^p)^2}\left(r_p^{-1} - \kappa^p \coth(\kappa^p r_p)\right) - \frac{K^m\left(r_p^{-1} + \kappa^m\right)}{(\kappa^m)^2}\right], \tag{40}$$

$$a_1 = \frac{\beta_{vA}^p}{D_v^p}\left(\frac{\alpha_i^m}{D_i^m} + r_p^{-1} + \kappa^m\right)\left[c_v^{p0} D_v^p\left(r_p^{-1} - \kappa^p \coth(\kappa^p r_p) - \frac{\alpha_i^p + \beta_{iA}^p}{D_i^p}\right) + \frac{K^p}{(\kappa^p)^2}\left(r_p^{-1} - \kappa^p \coth(\kappa^p r_p)\right)\right] +$$
$$\frac{\beta_{iA}^m}{D_i^m}\left(r_p^{-1} - \kappa^p \coth(\kappa^p r_p) - \frac{\alpha_i^p}{D_i^p}\right)\left(\frac{K^m\left(r_p^{-1} + \kappa^m\right)}{(\kappa^m)^2} - \beta_{vA}^m c_v^{m0}\right), \tag{41}$$

$$a_0 = \frac{\beta_{iA}^p}{D_i^p}\frac{K^p}{(\kappa^p)^2}\left(\kappa^p \coth(\kappa^p r_p) - r_p^{-1}\right)\left(\frac{\alpha_i^m}{D_i^m} + r_p^{-1} + \kappa^m + \frac{\beta_{vA}^m}{D_v^m}\right) +$$
$$\left[\beta_{vA}^m c_v^{m0}\left(\frac{\alpha_i^m}{D_i^m} + r_p^{-1} + \kappa^m\right) + \frac{\beta_{vA}^m}{D_v^m}\frac{K^m\left(r_p^{-1} + \kappa^m\right)}{(\kappa^m)^2}\right]\left(\kappa^p \coth(\kappa^p r_p) + \frac{\alpha_i^p + \beta_{iA}^p}{D_i^p} - r_p^{-1}\right). \tag{42}$$

If irradiation is absent $(K^m = 0, K^p = 0)$, the equilibrium impurity concentration at the boundary is as follows:

$$\frac{c_A^{m0}(r_p)}{c_A^p} = \frac{\beta_{vA}^m c_v^{m0}(r_p)}{\beta_{vA}^p c_v^{p0}(r_p)}. \tag{43}$$

## 3. Kinetics of evolution of the precipitate size

Let the impurity diffusion in the matrix is governed by the usual equation:

$$\frac{\partial c_A^m}{\partial t} = -\text{div } j_A^m, \quad j_A^m = -D_A^m \nabla c_A^m, \tag{44}$$

where $D_A^m$ is the impurity diffusion constant in the matrix.

Consider, that prior to irradiation the system precipitate-matrix was in its equilibrium state. Thus, the initial condition is

$$c_A^m(r,0) = c_A^{m0}, \tag{45}$$

where $c_A^{m0}$ is a thermodynamic equilibrium impurity concentration in the matrix.

The impurity flux transited across the boundary is given by the expression (21). In the first order in a deviation of the impurity concentration at the boundary from its equilibrium value (39), the expression (21) looks like:

$$j_A^m(r_p,t) = j_{At}(r_p,t) = \omega_0 \left[\beta_{iA}^m c_i^{meq}(r_p) + \beta_{vA}^p c_v^{peq}(r_p)\right]\left[c_A^{meq}(r_p) - c_A^m(r_p,t)\right]. \tag{46}$$

Consider, that in the matrix each precipitate occupies a spherical region with radius $R$, such, that the impurity flux at the boundary of this region is zero. Such a region is conventionally called an influence region. Thus, the second boundary condition looks like:

$$j_A^m(R,t) = 0. \tag{47}$$

Taking into account the initial (45) and the boundary conditions (46), (47), the impurity diffusion equation (44) has the solution:

$$c_A^m(r,t) = c_A^{m0} + \omega_0 \left[\beta_{iA}^m c_i^{meq}(r_p) + \beta_{vA}^p c_v^{peq}(r_p)\right]\left[c_A^{meq}(r_p) - c_A^{m0}\right]\frac{r_p}{r}\sum_{n=1}^{\infty}\frac{y_n(r)y_n(r_p)}{\lambda_n\|y_n\|^2}\cdot[1-\exp(-\lambda_n t)], \tag{48}$$

where

$$\|y_n\|^2 = \int_{r_p}^{R} y_n^2(x)dx. \tag{49}$$

Here the eigenfunctions of the Sturm-Liouville problem, corresponding to the given diffusion problem, are as follows:

$$y_n(x) = \sin\sqrt{\frac{\lambda_n}{D_A^m}}x + \tan\left[\arctan\left(\sqrt{\frac{\lambda_n}{D_A^m}}R\right) - \sqrt{\frac{\lambda_n}{D_A^m}}R\right]\cdot\cos\sqrt{\frac{\lambda_n}{D_A^m}}x, \tag{50}$$

and the corresponding eigenvalues $\lambda_n$ are roots of the equation

$$\arctan\left(\sqrt{\frac{\lambda_n}{D_A^m}}R\right) - \arctan\left[\sqrt{\frac{\lambda_n}{D_A^m}}\left(r_p^{-1} + \frac{\omega_0\left[\beta_{iA}^m c_i^{meq}(r_p) + \beta_{vA}^p c_v^{peq}(r_p)\right]}{D_A^m}\right)^{-1}\right] = \sqrt{\frac{\lambda_n}{D_A^m}}(R - r_p) - \pi n. \quad (51)$$

The velocity of change of the precipitate size is determined by the atomic flux transited across its boundary:

$$\frac{dr_p}{dt} = -\omega_0 j_t(r_p). \quad (52)$$

The atomic flux transited across the boundary is given by the expression (22). In the first order in a deviation of the impurity concentration at the boundary from its equilibrium value (39), the expression (22) looks like:

$$j_t(r_p, t) = \left[\beta_{iA}^m c_i^{meq}(r_p) + \beta_{vA}^p c_v^{peq}(r_p)\right]\left[c_A^{meq}(r_p) - c_A^m(r_p, t)\right] / c_A^p. \quad (53)$$

Taking into account the above-stated, the expression (52) gets the next explicit form:

$$\frac{dr_p}{dt} = \omega_0 \frac{\left[\beta_{iA}^m c_i^{meq}(r_p) + \beta_{vA}^p c_v^{peq}(r_p)\right]\left[c_A^{meq}(r_p) - c_A^{m0}\right]}{c_A^p} \times$$
$$\left\{\omega_0\left[\beta_{iA}^m c_i^{meq}(r_p) + \beta_{vA}^p c_v^{peq}(r_p)\right] \cdot \sum_{n=1}^{\infty} \frac{y_n^2(r_p)}{\lambda_n \|y_n\|^2} \cdot [1 - \exp(-\lambda_n t)] - 1\right\}. \quad (54)$$

The characteristic time for the impurity diffusion front to reach the boundary of the influence region is as follows:

$$\tau_R = (R - r_p)^2 / D_A^m. \quad (55)$$

In the case when the observation time is much less than $\tau_R$, one can consider the impurity diffusion to occur in an infinite matrix and the second boundary condition becomes:

$$c_A^m(\infty, t) = c_A^{m0}. \quad (56)$$

The solution of the equation (44) with the initial condition (45) and the boundary conditions (46), (56) has a more obvious look:

$$c_A^m(r, t) = c_A^{m0} + \frac{\omega_0\left[\beta_{iA}^m c_i^{meq}(r_p) + \beta_{vA}^p c_v^{peq}(r_p)\right]\left[c_A^{meq}(r_p) - c_A^{m0}\right]}{D_A^m\left(r_p^{-1} + \omega_0\left[\beta_{iA}^m c_i^{meq}(r_p) + \beta_{vA}^p c_v^{peq}(r_p)\right]/D_A^m\right)} \cdot \frac{r_p}{r}\left\{\text{erfc}\left(\frac{r - r_p}{2\sqrt{D_A^m t}}\right) - \right.$$
$$\exp\left[(r - r_p)\left(r_p^{-1} + \omega_0 \frac{\beta_{iA}^m c_i^{meq}(r_p) + \beta_{vA}^p c_v^{peq}(r_p)}{D_A^m}\right) + \left(r_p^{-1} + \omega_0 \frac{\beta_{iA}^m c_i^{meq}(r_p) + \beta_{vA}^p c_v^{peq}(r_p)}{D_A^m}\right)^2 D_A^m t\right] \times \quad (57)$$
$$\left.\text{erfc}\left(\frac{r - r_p}{2\sqrt{D_A^m t}} + \left(r_p^{-1} + \omega_0 \frac{\beta_{iA}^m c_i^{meq}(r_p) + \beta_{vA}^p c_v^{peq}(r_p)}{D_A^m}\right)\sqrt{D_A^m t}\right)\right\}.$$

At the same time the expression (54) also becomes simpler:

$$\frac{dr_p}{dt} = \omega_0 \frac{\left[\beta_{iA}^m c_i^{meq}(r_p) + \beta_{vA}^p c_v^{peq}(r_p)\right]\left[c_A^{meq}(r_p) - c_A^{m0}\right]}{c_A^p} \left\{-1 + \frac{\omega_0 \left[\beta_{iA}^m c_i^{meq}(r_p) + \beta_{vA}^p c_v^{peq}(r_p)\right]}{D_A^m \left(r_p^{-1} + \omega_0 \left[\beta_{iA}^m c_i^{meq}(r_p) + \beta_{vA}^p c_v^{peq}(r_p)\right]/D_A^m\right)} \times \right.$$
$$\left. \left[1 - \exp\left[\left(r_p^{-1} + \omega_0 \frac{\beta_{iA}^m c_i^{meq}(r_p) + \beta_{vA}^p c_v^{peq}(r_p)}{D_A^m}\right)^2 D_A^m t\right] \mathrm{erfc}\left[\left(r_p^{-1} + \omega_0 \frac{\beta_{iA}^m c_i^{meq}(r_p) + \beta_{vA}^p c_v^{peq}(r_p)}{D_A^m}\right)\sqrt{D_A^m t}\right]\right]\right\}. \quad (58)$$

It follows from the expression (58) that the kinetics of evolution of the precipitate radius has the characteristic time scale

$$\tau_{r_p} = \left(r_p^{-1} + \omega_0 \left[\beta_{iA}^m c_i^{meq}(r_p) + \beta_{vA}^p c_v^{peq}(r_p)\right]/D_A^m\right)^{-2} / D_A^m, \quad (59)$$

during which the impurity concentration at the boundary saturates to its equilibrium value $c_A^{meq}(r_p)$. At the initial stage of the precipitate evolution, when $t \ll \tau_{r_p}$, the velocity of dissolution is determined by the transboundary kinetics of migration of the impurity:

$$\frac{dr_p}{dt} \approx -\omega_0 \frac{\left[\beta_{iA}^m c_i^{meq}(r_p) + \beta_{vA}^p c_v^{peq}(r_p)\right]\left[c_A^{meq}(r_p) - c_A^{m0}\right]}{c_A^p}. \quad (60)$$

At the later stage of the precipitate evolution, when $t \gg \tau_{r_p}$, the velocity of dissolution is determined by the volume diffusion of the impurity

$$\frac{dr_p}{dt} \approx -\omega_0 \frac{\left[\beta_{iA}^m c_i^{meq}(r_p) + \beta_{vA}^p c_v^{peq}(r_p)\right]\left[c_A^{meq}(r_p) - c_A^{m0}\right]}{c_A^p \left(1 + r_p \omega_0 \left[\beta_{iA}^m c_i^{meq}(r_p) + \beta_{vA}^p c_v^{peq}(r_p)\right]/D_A^m\right)}. \quad (61)$$

## 4. Comparison of the model results to experimental data

To compare the results of this model to the experimental data [2] on dissolution of the MgO-based precipitates in the ferritic martensitic steel EM10 under electron irradiation with energy 1 MeV it is reasonable to make some additional simplifications. Since in the absence of irradiation both magnesium and oxygen are practically insolvable in the ferritic matrix [7], one can put

$$c_A^{m0}/c_A^{p0} = 0. \quad (62)$$

Since the melting temperature of MgO essentially exceeds that of a steel, in the absence of irradiation the equilibrium concentrations of vacancies should satisfy the inequality $c_v^{p0} \ll c_v^{m0}$. Then from (62), taking into account (43), it follows, that

$$\beta_{vA}^m = 0, \quad (63)$$

i.e. the flux of vacancies from the matrix into the precipitate is negligible.

Under the experimental conditions [2] considered, the rate of PD production both in the precipitate and in the matrix is of the order of $10^{-3}$ dpa/s, and the temperature is $T \leq 550°C$. Therefore, in the both phases the nonequilibrium vacancies produced by irradiation dominate over the equilibrium ones:

$$c_v^{p0} << K^p / (\kappa^p)^2 D_v^p, \tag{64}$$

$$c_v^{m0} << K^m / (\kappa^m)^2 D_v^m. \tag{65}$$

Therefore, in the expression for the impurity diffusion constant in the matrix we neglect the contribution from the equilibrium vacancies and leave only the contribution from the nonequilibrium PD, assuming lack of their recombination (see e.g. [1]):

$$D_A^m = 2 K^m \omega_0 / (\kappa^m)^2. \tag{66}$$

Since the maximal dose $\Phi_{max}^p$ reached for the precipitate (see Fig. 2 below) is much less than the dose $\Phi_R = K^p \omega_0 \tau_R \sim 100$ dpa, at which the diffusion front of the impurity reaches the boundary of the influence region with the radius $R \sim 10^3$ nm of the order of the radius of the focused electron beam, for comparison with the experimental data we use the expression (58).

Taking into account the assumptions (62) – (66) made, the results of numerical integration of the equation (58) with the initial conditions from Table 1 and the values of the model parametres from Table 2, together with the experimental data [2] on dissolution of the MgO-based precipitates in the EM10 steel under 1 MeV electron irradiation at several temperatures, are given in Fig. 2.

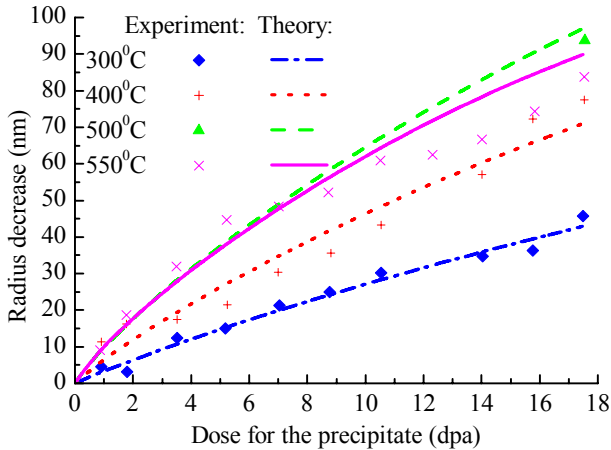

Fig. 2. Decrease of the precipitate radius vs. the dose absorbed by it. Points represent the experimental data [2]. Lines represent the numerical solution of the equation (58) with the initial conditions from Table 1 and the parametres from Table 2.

Table 1. The initial precipitate sizes

| Temperature, °C | 300 | 400 | 500 | 550 |
|---|---|---|---|---|
| Initial radius, nm | 100 | 125 | 150 | 120 |

Table 2. Values of the parametres entering the expression (58).

| $K^p \omega_0$, dpa/s | $1/\kappa^p$, nm | $K^m \omega_0$, dpa/s | $1/\kappa^m$, nm | $L$, nm | $r_d$, nm | $D_i^m / \beta_{iA}^m$, nm | $D_i^p / \beta_{iA}^p$, nm | $D_v^p / \beta_{vA}^p$, nm |
|---|---|---|---|---|---|---|---|---|
| $10^{-3}$ | 55 | $6 \cdot 10^{-3}$ | 10 | 10,5 | 0,6 | $0,3 \cdot \exp\left(\dfrac{2400}{T}\right)$ | $0,3 \cdot \exp\left(\dfrac{2400}{T}\right)$ | $3 \cdot \exp\left(\dfrac{3000}{T}\right)$ |

From Fig. 2 one can see, that at the chosen values of the model parametres the theory gives a good fit to the experimental data at low temperatures and a qualitative fit at 550ºC.

## 5. Discussion

The model of homogeneous semicoherent interphase boundary describes the processes of substance transport between a precipitate and a matrix alloy due to thermoactivated migration across the interphase boundary of irradiation-produced inequilibrium point defects. It allows to investigate the kinetics of substance redistribution and change of the precipitate size in alloys under irradiation.

In section 2 of the paper it is shown, that within the model irradiation leads to modification of the value of the impurity equilibrium concentration (39) in the matrix near the boundary. Expression (39) includes 11 parametres: $\alpha_n^\varphi / D_n^\varphi$, characterising an absorptive ability of the boundary and identical for all types of PD in both phases; 4 parametres $\beta_{nA}^\varphi / D_n^\varphi$, characterising a transparency of the boundary for PD from both sides, 2 parametres $\kappa^\varphi$, characterising a volume density of PD sinks in the phases, 2 parametres $K^\varphi$, characterising volume PD production rates under the influence of irradiation and 2 parametres $c_v^{\varphi 0}$, being thermodynamic equilibrium concentrations of vacancies in both phases.

In section 3 the kinetics of substance redistribution between the precipitate and the matrix is studied and the expression (54) for the velocity of the precipitate radius change under the influence of irradiation is found. The expression (54) and its subsequent simplifications include an additional parametre $D_A^m$, being the impurity diffusion constant in the matrix.

In section 4 the results of numerical solution of the equation for the velocity of the precipitate radius change under the influence of irradiation in the simplified form (58) are compared to the experimental data [2] on dissolution of the MgO-based precipitates in the ferritic ODS steel EM10 under 1 MeV electron irradiation. With the values of initial conditions from Table 1 and the model parametres from Table 2, the theory has a good agreement with the experiment at temperatures 300 ºC and 400 ºC and a qualitative agreement at 550 ºC.

**6. Summary**

- The model of homogeneous semicoherent interphase boundary provides a microscopic description for the processes of absorption and thermoactivated migration of irradiation-produced inequilibrium point defects at a semicoherent boundary between a heterophase precipitate and a substitution solid solution.
- The main result of the model is the expression for the atomic flux, transited across the boundary. According to the model, the flux increases with increasing the production rate of point defects and the temperature.
- The model allows a good fit to the experimental data [2] on dissolution of the MgO-based precipitates in the ferritic ODS steel EM10 under 1 MeV electron irradiation.

**Acknowledgements**

The author is grateful to Prof. A. Bakai, Dr. N. Lazarev and Dr. A. Turkin for introducing him into the physics of radiation effects in solids and numerous discussions.